\documentclass[12pt,fleqn]{article}
\usepackage{palatino}
\usepackage{graphicx}
\begin{document}
 
\leftline{\Large\bf Maximally informative dimensions:}
\leftline{\Large\bf  Analyzing neural responses to
natural signals}
\bigskip\bigskip

\leftline{\large Tatyana Sharpee,$^1$ Nicole C. Rust,$^2$ and William
Bialek$^{1,3}$}
\bigskip

\noindent $^1$ Sloan--Swartz Center for Theoretical Neurobiology and Department of
Physiology \\ University of California at San Francisco, San Francisco,
California 94143--0444\\
$^2$ Center for Neural Science, New York University, New York, NY 10003\\
$^3$ Department of Physics, Princeton University, Princeton, New Jersey 08544\\
{\it sharpee@phy.ucsf.edu, rust@cns.nyu.edu, wbialek@princeton.edu}
\bigskip\bigskip

\leftline{\today} 

\bigskip\bigskip\hrule\bigskip\bigskip

\begin{quote}  
  We propose a method that would allow for a rigorous statistical
  analysis of neural responses to natural stimuli, which are
  non--Gaussian and exhibit strong correlations.  We have in mind a
  model in which neurons are selective for a small number of stimulus
  dimensions out of the high dimensional stimulus space, but within
  this subspace the responses can be arbitrarily nonlinear. Existing
  analysis methods are based on correlation functions between stimuli
  and responses, but these methods are guaranteed to work only in the
  case of Gaussian stimulus ensembles. As an alternative to correlation
  functions, we maximize the mutual information between the neural
  responses and projections of the stimulus onto low dimensional
  subspaces. The procedure can be done
  iteratively by increasing the dimensionality of this subspace. Those
  dimensions that allow the recovery of all of the information between
  spikes and the full unprojected stimuli describe the relevant
  subspace. If the dimensionality of the relevant subspace indeed is
  small, it becomes feasible to map out the neuron's input--output
  function even under fully natural stimulus conditions. These ideas
  are illustrated in simulations on model visual neurons responding to
  natural scenes.
\end{quote}

\section{Introduction}
\label{sec:intro}
From olfaction to vision and audition, a growing number of experiments
\cite{Rieke95}--\cite{Ringach02} are examining the responses of
sensory neurons to natural stimuli.    Observing the full
dynamic range of neural responses may require using stimulus ensembles
which approximate those \cite{Rieke_book,Simoncelli01} 
occurring in nature, and it is an attractive
hypothesis that the neural representation of these natural signals may
be optimized in some way \cite{Barlow61}--\cite{Twer01}.
Many neurons exhibit  strongly nonlinear and adaptive responses that are unlikely to
be predicted from a combination of responses to simple
stimuli; in particular neurons have been shown to adapt to the distribution of sensory inputs, so that
any characterization of these responses will depend on context \cite{brenner00-adapt,Fairhall01}.
Finally, the variability of neural response decreases substantially when complex dynamical, rather than static,
stimuli are used \cite{MainenSejnowski}--\cite{canberra}. All of these
arguments point to the need for general tools to analyze the neural
responses to complex, naturalistic inputs.

The stimuli analyzed by sensory neurons are intrinsically high
dimensional, with dimensions $D \sim 10^2 - 10^3$. For example, in the
case of visual neurons, the input is specified as light intensity on a
grid of at least $10\times 10$ pixels. Each of the presented stimuli
can be described as a vector ${\bf s}$ in this high dimensional
stimulus space.  It is important that stimuli need not be pictured as
being drawn as isolated points from this space.  Thus, if stimuli are
varying continuously in time we can think of the stimulus ${\bf s}$ as
describing a recent window of the stimulus history (e.~g., the past
$K$ frames of the movie, with dimensionality $K$ times larger than for
the description of a single frame) and then the distribution of
stimuli $P({\bf s})$ is sampled along some meandering trajectory in
this space; we will assume this process is ergodic, so that we can
exchange averages over time with averages over the true distribution
as needed.

Even though direct exploration of a $D \sim 10^2 - 10^3$ dimensional
stimulus space is beyond the constraints of experimental data
collection, progress can be made provided we make certain assumptions
about how the response has been generated.  In the simplest model, the
probability of response can be described by one receptive field (RF)
or linear filter \cite{Rieke_book}.  The receptive field can be
thought of as a template or special direction ${\bf v}$ in the
stimulus space such that the neuron's response depends only on a
projection of a given stimulus ${\bf s}$ onto ${\bf v}$, although the
dependence of the response on this projection can be strongly
nonlinear. In this simple model, the reverse correlation method
\cite{Rieke_book,deBoer} can be used to recover the vector ${\bf v} $
by analyzing the neuron's responses to Gaussian white noise.  In a
more general case, the probability of the response depends on
projections $s_i= \hat e_i \cdot {\bf s}$ of the stimulus ${\bf s}$ on
a set of vectors $\{ \hat e_1, \,\hat e_2,\, ...\, ,\hat e_n \}$:
\begin{equation}
\label{ior}
P({\rm spike}|{\bf s})=P({\rm spike}) f(s_1,s_2, ...,s_n), 
\end{equation}
where $P({\rm spike}|{\bf s})$ is the probability of a spike given a
stimulus ${\bf s}$ and $P({\rm spike})$ is the average firing rate.
Even though the ideas developed below can be used to analyze
input--output functions $f$ with respect to different neural responses,
such as patterns of spikes in time \cite{stcov,B00a}, we choose a
single spike as the response of interest. The vectors $\{\hat e_i \}$
may also describe how the time dependence of stimulus ${\bf s}$ affects the
probability of a spike. We will call the subspace spanned by the
set of vectors $\{\hat e_i \}$ the relevant subspace (RS).

Equation (\ref{ior}) in itself is not yet a simplification if the
dimensionality $n$ of the RS is equal to the
dimensionality $D$ of the stimulus space.  In this paper we will use
the idea of dimensionality reduction \cite{brenner00-adapt,stcov,rob+bill-feature} and assume
that $n \ll D$. The input--output function $f$ in Eq.~(\ref{ior}) can
be strongly nonlinear, but it is presumed to depend only on a small
number of projections. This assumption 
appears to be less stringent than that of approximate linearity which
one makes when characterizing neuron's response in terms of Wiener
kernels (see, for example, the discussion in Section 2.1.3 of Ref. \cite{Rieke_book}). 
The most difficult part in reconstructing the input--output function is
to find the RS. Note that for $n>1$, a description in terms of
any linear combination of vectors $\{\hat e_i \}$ is just as valid,
since we did not make any assumptions as to a particular form of
nonlinear function $f$. 

Once the relevant subspace is known, the probability $P({\rm
  spike}|{\bf s})$ becomes a function of only few parameters, and it
becomes feasible to map this function experimentally, inverting the
probability distributions according to Bayes' rule:
\begin{equation}
\label{bayes}
 f(\{s_i\})=\frac{P(\{ s_i \}|{\rm spike})}{P(\{ s_i \} )}.
\end{equation}
If stimuli are chosen from a correlated Gaussian noise ensemble, then the neural response can
be characterized by the spike--triggered covariance method
\cite{brenner00-adapt,stcov,rob+bill-feature}. It can be shown that the dimensionality of the
RS is equal to the number of nonzero eigenvalues of a matrix given by
a difference between covariance matrices of all presented stimuli and
stimuli conditional on a spike. Moreover, the RS is spanned by the
eigenvectors associated with the nonzero eigenvalues multiplied by
the inverse of the {\it a priori} covariance matrix.  Compared to the
reverse correlation method, we are no longer limited to finding only
one of the relevant directions $\hat e_i$. Both the reverse
correlation and the spike--triggered covariance method, however, give
rigorously interpretable results {\it only} for Gaussian distributions
of inputs.

In this paper we investigate whether it is possible to lift the
requirement for stimuli to be Gaussian. When using natural stimuli,
which  certainly are non--Gaussian, the RS cannot be found by the
spike--triggered covariance method.  Similarly, the reverse correlation
method does not give the correct RF, even in the simplest case where
the input--output function in Eq. (\ref{ior}) depends only on one projection.
However, vectors that span the RS clearly are special
directions in the stimulus space independent of assumptions about $P({\bf s})$.  
This notion can be quantified by
Shannon information, and an optimization problem can be formulated to
find the RS.  We illustrate how the optimization scheme
works with natural stimuli for model orientation sensitive cells with
one and two relevant directions, much like simple and complex cells
found in primary visual cortex. It also is  possible to estimate
average errors in the reconstruction.  The advantage of this
optimization scheme is that it does not rely on any specific
statistical properties of the stimulus ensemble, and can thus be used with
natural stimuli.

\section{Information as an objective function} 
\label{sec:info}

When analyzing neural responses, we compare the {\it a priori} probability
distribution of all presented stimuli with the probability
distribution of stimuli which lead to a spike \cite{stcov}.  For Gaussian signals,
the probability distribution can be characterized by its second
moment, the covariance matrix. However, an ensemble of natural stimuli is
not Gaussian, so that neither second nor any other finite number of
moments is sufficient to describe the probability distribution.  In
this situation,  Shannon information provides the rigorous way of
comparing two probability distributions. The average information carried by
the arrival time of one spike is given by \cite{B00a}
\begin{equation}
\label{ispike}
I_{\rm spike}= \int d^D{\bf s} P({\bf s}|{\rm spike}) \log_2 \left[
{{P({\bf s}|{\rm spike})}\over
{P({\bf s})}}\right]\,.
\end{equation}
The information per spike as written in (\ref{ispike}) is difficult
to estimate experimentally, since it requires either sampling of the
high--dimensional probability distribution $P({\bf s}|{\rm spike})$ 
or a model of how spikes were generated, i.e. the knowledge of
low--dimensional RS. However it is possible to
calculate $I_{\rm spike}$ in a model--independent way, if stimuli are
presented multiple times to estimate the probability
distribution $ P({\rm spike}|{\bf s})$. Then,
\begin{equation}
\label{exp_ispike}
I_{\rm spike}=\left\langle \frac{P({\rm spike}|{\bf s})}{P({\rm spike})} \log_2\left[
\frac{P({\rm spike}|{\bf s})}{P({\rm spike})}\right]\right\rangle_{\bf s}, 
\end{equation}
where the average is taken over all presented stimuli.  As discussed
in \cite{B00a}, this is useful in practice because we can
replace the ensemble average $\langle \rangle_{\bf s}$ with a time
average, and $P({\rm spike}|{\bf s})$ with the time dependent spike rate
$r(t)$.  Note that for a finite dataset of $N$ trials, the obtained
value $I_{\rm spike}(N)$ will be on average larger than $ I_{\rm
  spike}(\infty)$, with difference $\sim N_{\rm stimuli}/(N_{\rm
  spike} \, 2 \ln 2)$, where $N_{\rm stimuli}$ is the number of
different stimuli, and $N_{\rm spike}$ is the number of elicited
spikes \cite{Treves}. The true value $I_{\rm spike}$ can also be found
by extrapolating to $N\to\infty$ \cite{B00a,entropy}. Measurement of
$I_{\rm spike}$ in this way provides a model independent benchmark
against which we can compare any description of the neuron's input--output relation.

Having in mind a model in which spikes are generated according to
projection onto a low dimensional subspace, we start by projecting all
of the presented stimuli on a particular direction ${\bf v}$ in the stimulus
space, and form probability distributions
\begin{eqnarray}
 P_{\bf v}(x|{\rm spike})&=&\langle\delta(x-{\bf s}\cdot{\bf v})|{\rm spike}\rangle_{\bf
  s},\\
  P_{\bf v}(x)&=&\langle\delta(x-{\bf s}\cdot {\bf v})\rangle_{\bf s},
\end{eqnarray}
where $\langle \cdots | {\rm spike}\rangle$ denotes an expectation value conditional on the occurrence of a
spike.  The information
\begin{equation}
\label{Iv}
I( {\bf v})=\int dx P_{\bf v}(x|{\rm spike})\log_2\left[{{P_{\bf v}(x|{\rm spike})}\over
{P_{\bf v}(x)}}\right]
\end{equation}
provides an invariant measure of how much the occurrence of a spike is
determined by projection on the direction ${\bf v}$.  It is a function
only of direction in the stimulus space and does not change when
vector ${\bf v}$ is multiplied by a constant.  This can be seen by
noting that for any probability distribution and any constant $c$,
$P_{c{\bf v}}(x)=c^{-1}P_{\bf v}(x/c)$.  When evaluated along any
vector, $I({\bf v})\leq I_{\rm spike}$.  The total information $I_{\rm
  spike}$ can be recovered along one particular direction only if
${\bf v}=\hat e_1$, and the RS is one dimensional.

By analogy with (\ref{Iv}), one could also calculate information $I({\bf v}_1,... ,{\bf v}_n)$
along a set of several directions $\{{\bf v}_1,... , {\bf v}_n \}$ based
on the multi-point probability distributions:
\begin{eqnarray}
P_{{\bf v}_1,...,{\bf v}_n}(\{x_i\}|{\rm spike})&=&
\Bigg\langle \prod_{i=1}^n\delta(x_i-{\bf s}\cdot {\bf v}_i) | {\rm spike} \Bigg\rangle_{\bf s}, \\
 P_{{\bf v}_1,...,{\bf v}_n}(\{x_i\})&=&\Bigg\langle\prod_{i=1}^n\delta(x_i-{\bf s}\cdot {\bf v}_i)\Bigg\rangle_{\bf s}.
\end{eqnarray}

If we are successful in finding all of the $n$ directions $\hat e_i$
in the input--output relation (\ref{ior}), then the information
evaluated in this subspace will be equal to the total information
$I_{\rm spike}$. When we calculate information along a set of $n$
vectors that are slightly off from the RS, the answer
is, of course, smaller than $I_{\rm spike}$ and is initially quadratic in
deviations $\delta  {\bf v}_i$.
One can therefore hope to find the RS by maximizing information
with respect to $n$ vectors simultaneously.  
The information does not
increase if more vectors outside the RS are included. 
For uncorrelated stimuli, any vector or a set of vectors that
maximizes $I({\bf v})$ belongs to the RS. 
On the other hand, the result of optimization
with respect to a number of vectors $k<n$ may deviate from the RS if stimuli
are correlated. 
To find the RS, we first
maximize $I({\bf v})$, and compare this maximum with $I_{\rm spike}$, which
is estimated according to (\ref{exp_ispike}). If the difference
exceeds that expected from finite sampling corrections, we increment
the number of directions with respect to which information is
simultaneously maximized.

The information $I({\bf v})$ as defined by (\ref{Iv}) is a continuous
function, whose gradient can be computed. We find
\begin{equation}
\label{grad}
\nabla_{\bf v} I=\int dx P_{\bf v}(x) \left[\langle {\bf s}|x, {\rm spike}\rangle-\langle {\bf s}|x\rangle \right]
\cdot \left[\frac{d}{dx}
\frac{P_{\bf v}(x|{\rm spike})}{P_{\bf v}(x)}\right],
\end{equation}
where 
\begin{equation}
\langle {\bf s}|x,{\rm spike}\rangle={1\over{P(x|{\rm spike})}}\int d^D{\bf s} \, {\bf s}\delta(x-{\bf s}\cdot
{\bf v})P({\bf
  s}|{\rm spike}),
\end{equation}
and similarly for $\langle {\bf s}|x\rangle$. Since information does
not change with the length of the vector, ${\bf v} \cdot \nabla_{\bf v}I =0 $ (which
can also be seen from (\ref{grad}) directly).

As an optimization algorithm, we have used a combination of gradient
ascent and simulated annealing algorithms: successive line
maximizations were done along the direction of the gradient.  During
line maximizations, a point with a smaller value of information was
accepted according to Boltzmann statistics, with probability $\propto
\exp[(I( {\bf v}_{i+1})-I( {\bf v}_i))/T]$.  The effective temperature $T$ is
reduced upon completion of each line maximization.

\section{Results}

We tested the scheme of looking for the most informative directions on
model neurons that respond to stimuli derived from natural scenes. As
stimuli we used patches of black and white photos digitized to 8 bits, in which
no corrections were made for camera's light intensity transformation
function.  Our goal is to demonstrate that even though the
correlations present in natural scenes are non--Gaussian, they can be
successfully removed from the estimate of vectors defining the RS.

\subsection{A model simple cell}

\begin{figure}[t]
\begin{center}
\includegraphics[width=3.4in]{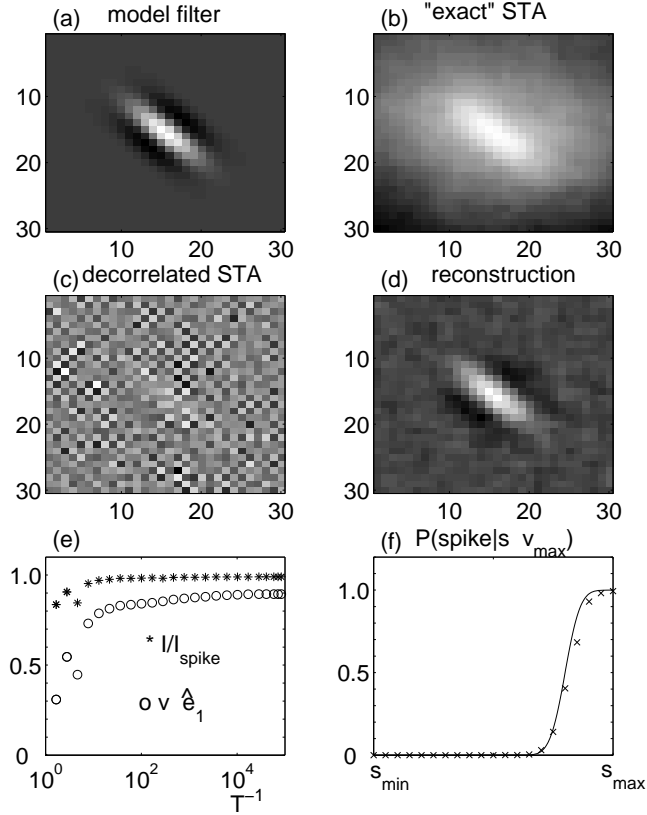}
\end{center}
\caption{Analysis of  a model 
  simple cell with RF shown in (a). The ``exact'' spike-triggered average $
  {\bf v}_{\rm\tiny sta}$ is shown in (b).  Panel (c) shows an attempt to
  remove correlations according to reverse correlation method, $C_{a\,
    priori}^{-1} {\bf v}_{\rm sta}$; (d) vector $ \hat v_{\rm max}$ found by
  maximizing information; (e) convergence of the algorithm according
  to information $I({\bf v})$ and projection $\hat v \cdot \hat e_1$ as a function of
  inverse effective temperature $T^{-1}$. (f)  The probability of a spike
  $P({\rm spike}|{\bf s} \cdot \hat v_{max})$ (crosses) is compared to
  $P({\rm spike}|s_1)$ used in generating spikes (solid line).
  Parameters $\sigma=0.05(s_{\rm max}-s_{\rm min})$ and $s_t=0.8(s_{\rm
    max}-s_{\rm min})$ [$s_{\rm max}$ and $s_{\rm min}$ are the
  maximum and minimum values of $s_1$ over the ensemble of presented
  stimuli].}
\label{fig:simple}
\end{figure}

Our first example is based on the properties of simple cells found
in the primary visual cortex.  A model phase and orientation sensitive
cell has a single relevant direction $\hat e_1$ shown in
Fig.~\ref{fig:simple}(a).  A given frame ${\bf s}$ leads to
a spike if the projection $s_1={\bf s} \cdot \hat e_1$ reaches a threshold value
$s_t$ in the presence of noise:
\begin{equation}
\label{model}
\frac{P({\rm spike}|{\bf s})}{P({\rm spike})}\equiv f(s_1)=\langle \theta ( s_1-s_t+\xi)\rangle,
\end{equation} 
where a Gaussian random variable $\xi$ of variance $\sigma^2$ models
additive noise, and the function $\theta(x)=1$ for $x>0$, and zero
otherwise. Together with the RF $\hat e_1$, the parameters
$s_t$ for threshold and the noise variance $\sigma^2$ determine the
input--output function.

The spike--triggered average (STA), or reverse correlation function
\cite{Rieke_book,deBoer}, shown in Fig.~\ref{fig:simple}(b), is
broadened because of spatial correlations present in the stimuli.  In
a model, the effect of noise on our estimate of the STA can be eliminated by
averaging the presented stimuli weighted with the exact firing rate,
as opposed to using a histogram of responses to estimate $P({\rm
  spike}|{\bf s})$ from a finite set of trials.  We have used this
``exact'' STA,
\begin{equation}
{\bf v}_{\rm sta} = \int d^D{\bf s}\, {\bf s} P({\bf s} | {\rm spike})
= {1\over{P({\rm spike})}}\int d^D {\bf s} P({\bf s})\, {\bf s} P({\rm spike}|{\bf s}) ,
\end{equation} 
in calculations presented in Fig.~\ref{fig:simple}(bc). If
stimuli were drawn from a Gaussian probability distribution, they
could be decorrelated by multiplying ${\bf v}_{\rm sta}$ by the
inverse of the {\it a priori} covariance matrix, according to the
reverse correlation method, ${\hat v}_{Gaussian\,est} \propto C_{a\,
  priori}^{-1} {\bf v}_{\rm sta}$. The procedure is not valid for
non--Gaussian stimuli and nonlinear input--output functions
(\ref{ior}).  The result of such a decorrelation is shown in
Fig.~\ref{fig:simple}(c).  It clearly is missing some of the structure
in the model filter, with projection $\hat e_1\cdot {\hat
  v}_{Gaussian\,est} \approx 0.14$. The discrepancy  is not due
to neural noise or finite sampling, since the ``exact'' STA was
decorrelated; the absence of noise in the exact STA also means that there would be
no justification for smoothing the results of the decorrelation.  
The discrepancy between the true receptive field and the decorrelated
STA increases with the strength
of nonlinearity in the input--output function.

In contrast, it is possible to obtain a good estimate of the relevant
direction $\hat e_1$ by maximizing information directly, see panel
(d).  A typical progress of the simulated annealing algorithm with
decreasing temperature $T$ is shown in Fig.~\ref{fig:simple}(e).
There we plot both the information along the vector, and its
projection on $\hat e_1$. The final value of projection depends on the
size of the data set, see below. In the example shown in
Fig.~\ref{fig:simple} there were $\approx 50,000$ spikes with average
probability of spike $\approx 0.05$ per frame, and the reconstructed
vector has projection ${\hat v}_{max}\cdot \hat e_1\approx 0.9$.
Having estimated the RF, one can proceed to sample the nonlinear
input-output function. This is done by constructing histograms for
$P({\bf s} \cdot {\hat v}_{\rm max})$ and $P({\bf s} \cdot {\hat v}_{\rm
  max}|{\rm spike})$ of projections onto vector $\hat v_{\rm max}$
found by maximizing information, and taking their ratio, as in
Eq.~(\ref{bayes}). In Fig.~\ref{fig:simple}(f) we compare $P({\rm
  spike}|{\bf s} \cdot {\hat v}_{\rm max})$ (crosses) with the
probability $P({\rm spike}|s_1)$ used in the model (solid line).

\subsection{Estimated deviation from the optimal direction}

When information is calculated from  a finite data set, the
vector ${\bf v}$ which maximizes $I$ will deviate from the true RF
$\hat e_1$. The deviation $\delta {\bf v}={\bf v}-\hat e_1$ arises because
the probability distributions are estimated from experimental
histograms and differ from the distributions found in the limit on
infinite data size. For a simple cell, the quality of reconstruction
can be characterized by the projection ${\bf v} \cdot \hat e_1=1- \frac{1}{2}
\delta {\bf v}^2$, where both ${\bf v}$ and $\hat e_1$ are normalized, and
$\delta {\bf v}$ is by definition orthogonal to $\hat e_1$.  The deviation
$\delta {\bf v} \sim A^{-1} \nabla I$, where $A$ is the Hessian of
information. Its structure is similar to that of a covariance matrix:
\begin{equation}
A_{ij}=\frac{1}{\ln 2}\int dx P(x|{\rm spike})\left (\frac{d}{dx} \ln \frac{P(x|{\rm spike})}{P(x)}\right)^2( \langle s_is_j|x\rangle-\langle s_i|x\rangle \langle s_j|x\rangle).
\end{equation}

\begin{figure}[t]
\begin{center}
\includegraphics[width=3.2in]{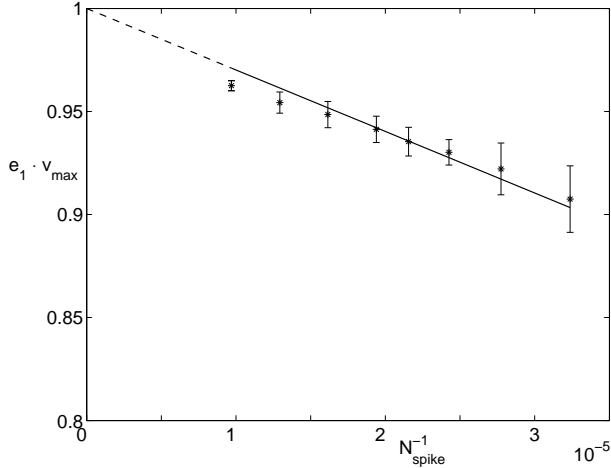}
\caption{Projection of vector $\hat v_{\rm max}$ that maximizes information
  on RF $\hat e_1$ is plotted as a function of the number of spikes to
  show the linear scaling in $1/N_{\rm spike}$. In this series of
  simulations, the average probability of a spike
  (\protect\ref{model}) had parameter values
  $\sigma=0.1(s_{\rm max}-s_{\rm min})$ and $s_t=0.6(s_{\rm
    max}-s_{\rm min})$.}
\label{fig:scaling}
\end{center}
\end{figure}

When averaged over possible outcomes of $N$ trials, the gradient of
information is zero for the optimal direction. Here in order to
evaluate $\langle \delta {\bf v}^2\rangle={\rm Tr}[A^{-1}\langle \nabla I
\nabla I^{T}\rangle A^{-1}] $, we need to know the variance of the gradient
of $I$.  By discretizing both the space of
stimuli and possible projections $x$, and assuming that the
probability of generating a spike is independent for different bins,
we estimate $\langle \nabla I_i \nabla I_j \rangle\sim
A_{ij}/( N_{\rm spike}\ln 2)$. Therefore an expected error in the
reconstruction of the optimal filter is inversely proportional to the
number of spikes and is given by:
\begin{equation}
\label{error}
1-{\bf v} \cdot \hat e_1 \approx \frac{1}{2}\langle \delta {\bf v}^2\rangle= \frac{{\rm Tr} [ A^{-1}] }{2 N_{\rm
spike}\ln 2} 
\end{equation}
In Fig.~\ref{fig:scaling} we plot the average projection of the normalized 
reconstructed vector ${\bf v}$ on the RF $\hat e_1$, and show that it scales correctly
with the number of spikes. 

\subsection{A model complex cell}

\begin{figure}[ht]
\begin{center}
\includegraphics[width=3.4in]{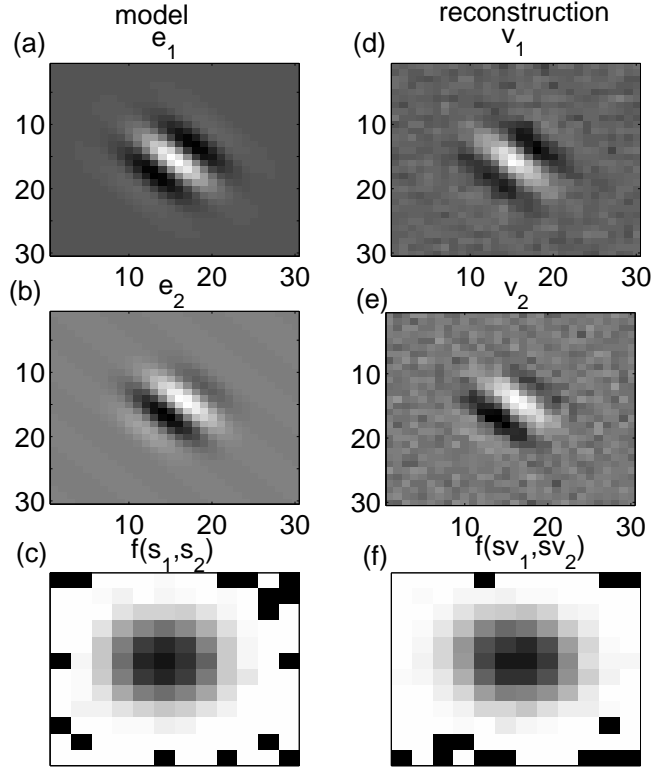}
\end{center}
\caption{Analysis of a model complex cell with relevant direction $\hat e_1$ and
  $\hat e_2$ shown in (a) and (b). Spikes are generated according to
  an ``OR'' input-output function $f(s_1,s_2)$ with the threshold
  $s_t=0.6(s_{\rm max}-s_{\rm min})$ and noise variance
  $\sigma=0.05(s_{\rm max}-s_{\rm min})$. Panel (c) shows how the
  input-output function is sampled by our ensemble of stimuli. Dark
  pixels for large values of $s_1$ and $s_2$ correspond to cases where
  $P(s_1,s_2)=0$. On the right, we show vectors ${\bf v}_1$ and ${\bf v}_2$ found
  by maximizing information $I({\bf v}_1,{\bf v}_2)$ together with the
  corresponding input-output function with respect to projections
  ${\bf s} \cdot {\bf v}_1$ and ${\bf s} \cdot {\bf v}_2$, panel (f). }
\label{fig:complex}
\end{figure}

A sequence of spikes from a model cell with two relevant directions
was simulated by projecting each of the stimuli on vectors that differ
by $\pi/2$ in their spatial phase, taken to mimic properties of
complex cells, as in Fig.~\ref{fig:complex}. A particular frame leads to
a spike according to a logical OR, that is if either $s_1={\bf s} \cdot \hat
e_1$, $-s_1$, $s_2={\bf s} \cdot \hat e_2$, or $-s_2$ exceeds a threshold
value $s_t$ in the presence of noise. Similarly to (\ref{model}),
\begin{equation}
\frac{P({\rm spike}| {\bf s})}{P({\rm spike})}=f(s_1,s_2)=\langle
\theta(|s_1|-s_t-\xi_1) \;\vee\;
\theta(|s_2|-s_t-\xi_2)\rangle\,,
\end{equation}
where $\xi_1$ and $\xi_2$ are independent Gaussian variables.
The sampling of this input--output function by our particular set of
natural stimuli is shown in Fig.~\ref{fig:complex}(c). Some,
especially large, combinations of values of $s_1$ and $s_2$ are not
present in the ensemble.
As is well known, reverse correlation fails in this case because the spike--triggered
average stimulus is zero, although with Gaussian stimuli the spike--triggered covariance
method would recover the relevant dimensions.  Here we show that searching for maximally
informative dimensions allows us to recover the relevant subspace even under more natural stimulus conditions.

We start by maximizing information with respect to one direction.
Contrary to the result Fig.~\ref{fig:simple}(e) for a simple cell, one
optimal direction recovers only about 60\% of the total information
per spike [Eq.~(\ref{exp_ispike})].  Perhaps surprisingly, because of
the strong correlations in natural scenes, even projection onto a
random vector in the $D\sim 10^3$ dimensional stimulus space has a high
probability of explaining 60\% of total information per spike.  We
therefore go on to maximize information with respect to two
directions.
An example of the reconstruction of input--output function of a complex
cell is given in Fig.~\ref{fig:complex}. Vectors ${\bf v}_1$ and ${\bf v}_2$ that
maximize $I({\bf v}_1,{\bf v}_2)$ are not orthogonal, and are also rotated with
respect to $\hat e_1$ and $\hat e_2$. However, the quality of reconstruction
 is independent of a particular choice of basis with the RS. The appropriate measure of
similarity between the two planes is the dot product of their
normals. In the example of Fig.~\ref{fig:complex},   $\hat n_{(\hat e_1,\hat e_2)}\cdot \hat n_{({\bf v}_1,{\bf
v}_2)}\approx 0.8$.

Maximizing information with respect to two directions requires a
significantly slower cooling rate, and consequently longer
computational times. However, the expected error in the reconstruction,
$1-\hat n_{(\hat e_1,\hat e_2)}\cdot \hat n_{({\bf v}_1,{\bf v}_2)}$,
follows a $N^{-1}_{\rm spike}$ behavior, similarly to (\ref{error}),
and is roughly twice that for a simple cell given the same number of
spikes.

\section{Remarks}
 
In conclusion, features of the stimulus that are most relevant for
generating the response of a neuron can be found by maximizing
information between the sequence of responses and the projection of
stimuli on trial vectors within the stimulus space. Calculated in this
manner, information becomes a function of direction in a stimulus
space. Those directions that maximize the information and account for
the total information per response of interest span the relevant
subspace.  This analysis allows the reconstruction of the relevant
subspace without assuming a particular form of the input--output
function. It can be strongly nonlinear within the relevant subspace,
and is to be estimated from experimental histograms.  Most
importantly, this method can be used with any stimulus ensemble, even
those that are strongly non--Gaussian as in the case of natural images.

\section*{Acknowledgments}
 We thank K.~D.~Miller for many helpful discussions. Work at UCSF was
supported in part by the Sloan and Swartz Foundations and by a training
grant from the NIH. Our collaboration began at the Marine
Biological Laboratory in a course supported by grants from NIMH and the Howard
Hughes Medical Institute.


\end{document}